\documentclass[12pt,preprint]{aastex}

\newcommand{\nix}{$\cdot\cdot\cdot$}

\slugcomment{To appear in ApJ Letters}

\shorttitle{X-ray core of NGC 253}
\shortauthors{Weaver, Heckman, Strickland \& Dahlem}

\begin{document}


\title{Chandra Observations of the Evolving Core of  
the Starburst Galaxy NGC 253}


\author{K. A. Weaver\altaffilmark{1}}
\affil{Laboratory for High Energy Astrophysics,
NASA/Goddard Space Flight Center, Greenbelt, MD 20771, USA}
\email{kweaver@milkyway.gsfc.nasa.gov}

\author{T. M. Heckman and D. K. Strickland\altaffilmark{2}}
\affil{Johns Hopkins University, Department of Physics and
Astronomy, Homewood campus, 3400 North Charles Street,
Baltimore, MD 21218, USA}

\author{M. Dahlem\altaffilmark{3}}
\affil{European Southern Observatory, Casilla 19001
Santiago 19, CHILE}


\altaffiltext{1}{also Johns Hopkins University}
\altaffiltext{2}{Chandra Fellow}

\begin{abstract}

{\it Chandra} observations of the core of the nearby starburst galaxy
NGC~253 reveal a heavily absorbed source of hard X-rays embedded within
the nuclear starburst region.   The source has an 
unabsorbed, 2 to 10 keV luminosity of $\ge10^{39}$ erg s$^{-1}$
and photoionizes the surrounding gas.
We observe this source through a dusty torus with a 
neutral absorbing column density of $N_{\rm H}\sim2\times10^{23}$
cm$^{-2}$.  The torus is hundreds of pc across and collimates the 
starburst-driven nuclear outflow.   
We suggest that the ionizing source is an intermediate-mass 
black hole or a weakly accreting supermassive black hole, which  
may signal the beginnings or endings of AGN activity.

\end{abstract}


\section{Introduction}

In recent years, there has been increasing speculation about 
the connection between circumnuclear starbursts and 
active galactic nuclei (AGN).   Such speculation is due
mostly to the fact that, as our instruments allow us to probe
closer to the cores of nearby galaxies, we find that an 
an increasing number contain starbursts and AGN in close
proximity (Levenson, Weaver and Heckman
2001, and references therein).  It is not clear whether proximity 
implies a physical connection, but a circumnuclear starburst
could easily provide a pathway toward forming a supermassive black 
hole (and subsequent AGN), since it can 
processes as much as $\sim10^{10}$ M$_{\sun}$
of material in $10^7 - 10^8$ years (Norman and Scoville 1988).
To date, however, there has been little direct evidence 
for this scenario. 

X-rays can penetrate the dense 
cores of nearby galaxies and are thus crucial for 
probing the possible links between starburst and AGN activity.
For this purpose, we have obtained {\it Chandra} observations  
of the nearby ($\sim2.6$ Mpc) starburst galaxy NGC 253.
This galaxy possesses a strong circumnuclear starburst  
(Strickland et al. 2000) and
evidence for a weak AGN (Turner \& Ho 1995,
Mohan, Anantharamaiah \& Goss 2002).
{\it Chandra}, with its resolution of 
$\sim1^{\prime\prime}$ ($\sim12$ pc at NGC~253) allows us to 
untangle the X-ray emission processes due to stellar 
and non-stellar activity for the first time at the 
core of the galaxy.
 
\section{Data Analysis}

The {\it Chandra} data were obtained with the 
CCD Imaging Spectrometer (ACIS) on 1999 December 16
(see Strickland et al.\ 2000).  The absolute astrometric
accuracy of the nuclear pointing is $<1^{\prime\prime}$ rms.
We used CIAO version 2.1.3 and HEASOFT version 5.1 for 
data analysis.
In addition, we reprocessed the data with gain-correction
files optimized for the actual focal-plane temperature 
of $-120^{\circ}$ C (acisD2000...gainN0003.fits),
as opposed to $-110^{\circ}$ C (Strickland et al.\ 2000).  

The X-ray spectrum of the galaxy core contains $\sim580$ 
photons, which is adequate for fitting spectral models 
if the data are binned as a function of energy. 
On the other hand, the narrow spectral features are 
better seen in the unbinned data.
We therefore examine the binned and the unbinned 
spectrum, using the statistical 
techniques of $\chi^2$ and the c-statistic, respectively.

\section{The ACIS Image}

A three color composite {\it Chandra} image of the central 
$1.7^{\prime} \times 2^{\prime}$  
(1.4 x 1.6 kpc) region of NGC 253 is shown in Figure 1.  The 
cross marks the position of the compact radio nucleus (Turner and Ho 1985),
which is assumed to lie at the core of the galaxy.  
The colors have been optimized with energy to best show the 
hard X-ray, arc-shaped feature (seen in green) that lies along 
the plane of the galaxy.  The arc
is approximately $5^{\prime\prime}$ by $23^{\prime\prime}$ (60 
pc by 300 pc); its color indicates significant foreground 
absorption of soft X-rays.

Figure 2 shows a closer view 
(26$^{\prime\prime}$ by 37$^{\prime\prime}$) in the three
energy bands that make up Figure 1.   The solid green line
marks the location of an extended region of radio emission, 
which is interpreted as a $\sim60\times300$ pc rotating,
dusty torus (Israel, White and Baas 1995, Dahlem et al. in 
prep.).  The X-ray spectrum of the galaxy core indicates 
$N_{\rm H} = 10^{23}$ cm$^{-2}$ through the torus, which is 
consistent with the H$_2$ column densities
deduced from the interferometer maps of this region 
(Frayer, Seaquist and Frail 1998, Peng et al.\ 1996).
The position of the arc with respect to the 
outflow suggests that the torus is the collimating 
mechanism for the outflow, and so we will refer to it 
throughout the rest of the paper as the collimating torus.
The inner edge of the collimating torus
is defined by a loop of optical/IR sources
while OH maser and compact radio sources mark the nuclear ridge
of star formation in the dense regions
where the gas piles up as it orbits in the bar 
potential (Peng et al.\ 1996).  

The hard X-ray emission is partly extended along the nuclear 
ridge, coincident with the radio and maser sources and 
confined to a relatively small, 5$^{\prime\prime}$ 
($\sim50$ pc) area.  It is not clear whether the peak in
hard X-ray emission is point like, but it 
is coincident with the compact synchrotron 
source at the center of the galaxy, which has  
a flat radio spectrum ($\alpha = 0.04\pm0.06$) 
and a brightness temperature of T$_b \sim 10^5$ K.
This temperature is too high to be due to free-free emission from H II
regions, but is within the upper limit for the brightness 
temperature in starbursts (Condon et al.\ 1991).  On the other
hand, hints of a non-starburst origin are found from 
observations of radio recombination line emission, which  
implies a source of ionizing radiation with a flux
of $6-20\times10^{51}$ photons s$^{-1}$
(Mohan, Anantharamaiah \& Goss (2002). 

We note that the {\it Chandra} data are not sensitive to the hard X-ray
emission on much larger scales, such as that observed with XMM
(Pietsch, et al.\ 2000).

\section{The Nuclear Spectrum}

The nuclear spectrum is extracted from within a circular 
region of diameter $5^{\prime\prime}$ centered $1^{\prime\prime}$ to the 
northeast of the radio nucleus (to avoid contamination from 
a young star cluster associated with the peak in the medium 
X-ray image in Figure 2b).  
The results of Gaussian fits to the
emission lines are listed in Table 1.  All lines are 
detected with a confidence level of $\ge90\%$  
according to an F-test.  The 
core spectrum, binned for illustration purposes,
is shown in Figure 3a. 

He-like triplets of Mg, Si and S indicate a strong contribution
from an optically thin plasma.  However, the 
line equivalent widths are unusually large and
inconsistent with a thermal gas with solar abundance ratios
for a plasma in ionization equilibrium
(Mewe et al. 1985; Kaastra 1992; Liedahl, Osterheld and Goldstein 1995).
This is borne out by the binned spectrum, which cannot 
be fitted with single or multi-temperature thermal plasma models with 
solar abundances (Models 1, 2, and 3, Table 2).  
When the high-temperature Mekal component is replaced with
a power law (Model 4), an excellent fit is obtained 
($\chi^2_{\nu}$=1.03).  The problem is that the photon
index of $-0.3$ is much smaller than observed for any 
known type of X-ray emitter.  Allowing a second, larger
absorbing column in front of the high-energy component,
results in a more realistic description of    
either a thermal plasma with solar abundances (Model 5) or a 
power law with reasonable values of $\Gamma=1.4-1.9$,
as might be expected for an AGN or XRBs (Model 6). 
For Model 6 the best-fitting absorbing column density 
is $N_{\rm H}\sim2\times10^{23}$ cm$^{-2}$
and the unabsorbed $2-10$ keV luminosity 
is $2\times10^{39}$ erg s$^{-1}$.

Although the binned spectrum is statistically well described by 
models 5 or 6, the features between 3 and 6 keV that show up 
prominently in the unbinned data are not predicted by either model.
We have verified that similar features exist in the archived XMM data. 
Such features are presumably due to Ar and Ca, and are predicted
at or near their observed energies for a highly  
photoionized plasma (Bautista \& Kallman 2000). 
We therefore added another component to our model; 
this one representing emission from a spherical distribution 
of clouds with density $10^9$ cm$^{-3}$ that 
fully cover a central, power-law source with $\Gamma\sim1.9$.
The model was generated with {\it xstar}
(heasarc.gsfc.nasa.gov/docs/software/xstar), and 
represents conditions similar to the core of a 
Seyfert 1 galaxy (T. Yaqoob, private communication).
The final, best-fitting model is a three-component hybrid 
plasma model, consisting of a moderately-absorbed, low-temperature
Mekal plasma plus a heavily-absorbed power law and emission  
from a photoionized plasma with ionization parameter 
$\xi \sim700$ (Figure 3b).  The results from fitting this 
model to the binned spectrum are listed in Table 2 (Model 7).

\section{The Nature of the Hard X-ray Source}

Our analysis of the 
{\it Chandra} data suggests the presence of a 
significant source of photoionization at the center of 
the nearby galaxy NGC~253.
This result compliments the recent discovery of resolved
(FWHM$\sim200$ km s$^{-1}$), radio recombination line emission 
at the same location by Mohan et al. (2002).  These authors 
rule out a compact SNR and young star cluster as the ionizing
source. Similarly, we find such sources implausible because they 
would produce a softer ionizing X-ray spectrum than observed.
We therefore consider other plausible sources of hard X-rays.

\subsection{Inverse-Compton Emission}

We can estimate the contribution of inverse Compton emission in NGC 253
following the arguments in Moran, Lehnert, \& Helfand (1999).
The ratio of the inverse Compton and synchrotron luminosities
($L_{IVC}/L_{S}$) is given
by the ratio of the energy density in the seed photon field to the energy
density in the magnetic field.
To estimate the former, we first use
the well-known radio-infrared correlation for star-forming galaxies
(Condon 1992) and the high resolution 20-cm VLA radio map from Ulvestad
\& Antonucci (1997) to deduce that the far-infrared luminosity
of the central-most 5 arcsec in NGC 253 is $5 \times 10^{42}$ erg s$^{-1}$.
Adopting the mean bolometric correction of 1.75 measured for dusty starburst
galaxies by Calzetti et al. (2000), the bolometric surface-brightness
of the nucleus of NGC 253 is then $B_{bol}$ = 320 erg cm$^{-2}$ s$^{-1}$
and the radiant energy density is $U_{rad} \sim B_{bol}/c = 1.1 \times 10^{-8}$
erg cm$^{-3}$. To estimate the energy density in the magnetic field, we apply
standard minimum-energy assumptions
to the VLA 20cm radio data. We take Condon's (1992) equation 13 with
$\beta$ = 40 (the galactic value for the proton/electron energy ratio)
and a path length of 60 pc (5 arcsec). This implies a magnetic field
strength of 270 $\mu$G and an energy density
$U_B = B^2/8\pi = 2.8 \times 10^{-9}$ erg cm$^{-3}$.

Thus, we estimate that $L_{IVC}/L_{S}$
= 3.8. At 20 cm, the monochromatic power of the central 5 arcsec
is $\nu$P$_{\nu}$ = 7.5 $\times 10^{36}$ erg s$^{-1}$, while the
corresponding power at 4.5 keV is $1.2 \times 10^{39}$ erg s$^{-1}$.
This suggests that the inverse Compton process makes only a minor
contribution to the hard X-ray emission. However, we emphasize that
our estimate is rough.  Within the context of our estimate, a
dominant inverse Compton contribution to the hard X-rays seemingly
requires that the magnetic field strength is about a factor
of 7 below the minimum-energy value.

\subsection{An Evolved Starburst}

The nuclear starburst in NGC~253 is approximately 
20 to 30 million years old (Engelbracht et al.\ 1998),
which implies that its hard X-ray emission is 
dominated by XRBs (Van Bever \& Vanbeveren 2000). 
Only a handful of bright XRBs would be required to produce the
observed $2-10$ keV luminosity of $2\times10^{39}$ erg s$^{-1}$
(Grimm, Gilfanov \& Sunyaev 2001).
On the other hand, the equivalent widths of the emission lines 
are tens to hundreds of times larger than those of Galactic XRBs
(Asai et al.\ 2000).  If several bright XRBs are the 
source of the continuum photons, they
must be located behind a significant amount of absorbing 
material that suppresses the photoionizing continuum.

We can estimate the continuum luminosity required 
to produce the observed line strengths. 
For an ionization parameter of $\sim700$ and a 
column density of $10^{23}$ cm$^{-2}$, line luminosities of 
$\sim10^{37}$ erg s$^{-1}$ require a 
a continuum luminosity of at least $\sim10^{40}$ erg s$^{-1}$
(Kallman 1991).  This would make the continuum source an 
ultraluminous compact X-ray source\footnote{Having an X-ray
luminosity higher than $10^{38}$ erg s$^{-1}$,
the Eddington luminosity of a 1.4 solar mass accreting neutron star.}
and similar to the luminous X-ray source near 
the center of M~82 (Matsumoto et al.\ 2001).   
Given the location of the ultraluminous source 
coincident with the radio core of NGC~253, chances
are high that it is an 
intermediate mass, accreting black hole (IMBH).
 
\subsection{A Buried AGN}

If an IMBH could be lurking at the core of NGC~253, then might 
a low-luminosity AGN (LLAGN) be lurking there as well?  Interestingly,
Mohan et al. (2002) find that, if they assume an AGN is responsible 
for producing the radio recombination line emission, the observed 
X-ray luminosity is $\sim1,000$ times less than that required to produce 
the ionizing photon flux.  This deficiency in X-ray flux 
is easily explained if, in addition
to intersecting the collimating torus, our line of sight 
to the central source is blocked by gas that
is Compton thick, with column density of $\ge10^{24}$ cm$^{-2}$.
If the hard X-ray power-law continuum represents
scattered X-rays and NGC~253 is 
similar to Seyfert 2 galaxies with buried nuclei (Awaki et al. 2000),
this would make the intrinsic X-ray luminosity at   
least $\sim10^{41}$ erg s$^{-1}$.  The relatively low luminosity
compared to other AGN might then result from advection dominated
accretion or the lack of sufficient fuel, which is
possible if the AGN is turning on or off.

Our proposed geometry for the core of NGC~253 is shown in Figure 4. 
The central source, possibly an IMBH or LLAGN, sits 
at the kinematic center of
the $\sim300$ pc diameter collimating torus.  The kinetic energy 
of the starburst has forced gas that used to occupy this cavity
into the base of the starburst-driven wind, 
clearing a path for X-rays to escape.  
The X-rays are scattered from and also photoionize 
the surrounding material.  Additional obscuration is 
provided by a starburst ring that provides
the radiation force to support the torus (Ohsuga \& Umemura 1999).
This material may be Compton thick.  The high ionization
parameter of the surrounding gas suggests physical 
conditions that are similar to ionized 
absorbers in Seyfert 1 galaxies, but in this case we would be 
seeing the region from the side, in emission,
rather than face on, in absorption.

We have shown that 
conditions at the center of NGC~253 provide fertile 
ground for studying the connection between starburst and 
AGN activity.  The gas associated with the circumnuclear starburst 
is responsible for obscuring the central continuum source.
This scenario has been predicted for the early stages of AGN
formation (e.g., Sanders et al.\ 1988), but more importantly, 
it shows that the putative $\sim$pc-scale molecular torus
in Seyfert 2 galaxies is not the only mechanism by which an
AGN-like continuum source can be hidden from our view.  
We speculate that NGC~253 is in an evolutionary
state where it is transitioning between a starburst 
and AGN phase.

\section{Conclusions}

{\it Chandra} X-ray observations of the nearby starburst galaxy
NGC~253 reveal what may be the beginnings or endings of AGN activity.
The excellent spatial resolution allows us to isolate the 
optically-thick torus that collimates the starburst-driven nuclear outflow. 
At the center of the torus, along with the evolved, circumnuclear starburst,
is a source of hard X-rays with an
unabsorbed, 2 to 10 keV luminosity of $\ge10^{39}$ erg s$^{-1}$.
We suggest that this ionizing source is an intermediate-mass
black hole or weakly accreting supermassive black hole.
These data provide a unique look at the 
complex interplay between starburst and AGN activity. 

Future tests of the starburst-AGN scenario will require studying
older starburst populations in nearby galaxies.  In particular,
it is important to look at normal stars in the optical or at  
X-ray binaries with {\it Chandra} in the more evolved, more AGN-like
composite (Seyfert/starburst) galaxies.

\bigskip
\acknowledgments
These observations were made using the Chandra X-ray Observatory, operated 
by NASA and the Harvard-Smithsonian Astrophysical Observatory.  The authors 
thank Tahir Yaqoob for supplying the photoionization model and the referee,
Joris Van Bever, for helpful suggestions.

\clearpage

\begin{figure}
\epsscale{0.7}
\plotone{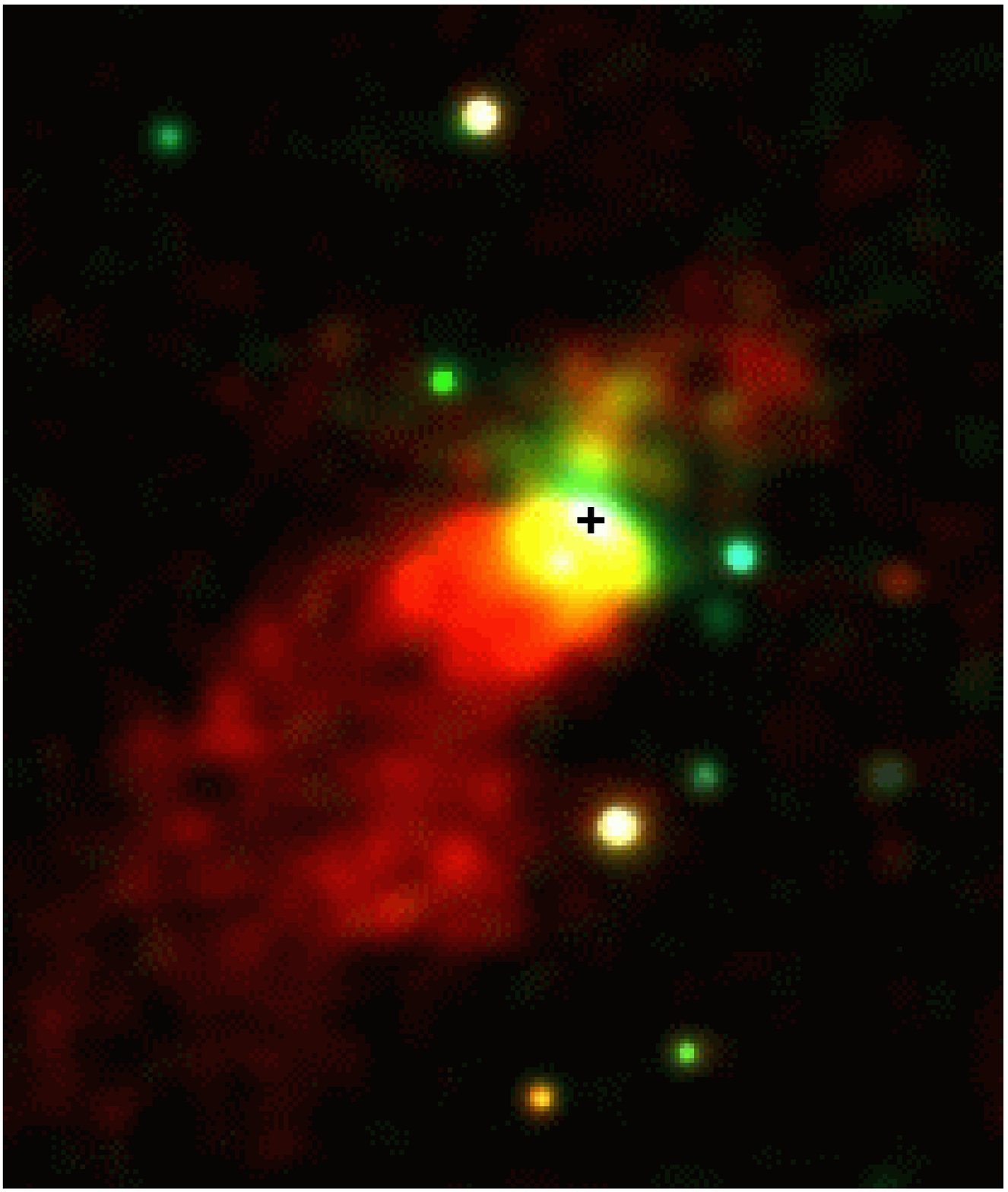}
\caption{Three color composite Chandra X-ray image
of the central 1.75$^{\prime}$ by 2$^{\prime}$ 
(1.4 kpc x 1.6 kpc) region of NGC~253,
smoothed to achieve a local S/N of 3.
Red, yellow and blue indicate the X-ray ``colors''
of 0.2--1.5 keV (soft), 1.5--4.5 keV (medium)
and 4.5--8 keV (hard),
respectively.  
The cross marks the position of the radio core (Turner and Ho 1985).
\label{fig1}
}
\end{figure}

\clearpage

\begin{figure}
\epsscale{1.0}
\plotone{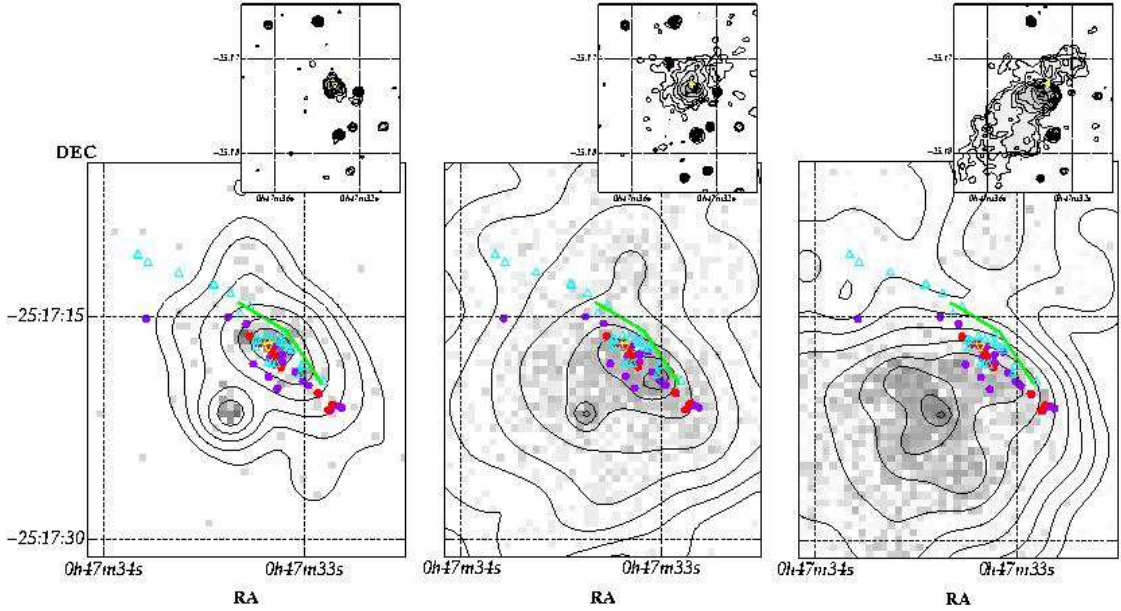}
\caption{
Hard, medium and soft X-ray images (the three bands
that are combined to produce Figure 1) with data from 
other wavebands.  Grey-scale represents the unbinned (0.5$^{\prime\prime}$ x
0.5$^{\prime\prime}$) and unsmoothed
{\it Chandra} image.  The thin black lines represent the brightness
contours of the adaptively smoothed images.
Contours are logarithmic and start at 2$\sigma$ above the 
{\it local} background;  the soft contours range from 0.1 to 14.0
counts/pixel; medium contours range from 0.04 to 21.0 counts/pixel;
hard contours range from 0.03 to 11.0 counts/pixel. 
Purple circles mark the compact radio sources
(Ulvestad and Antonucci 1997).  Red circles represent
compact OH maser features (Frayer, Seaquist and Frail 1998).
Blue triangles mark discrete optical and IR sources (Forbes et al. 2000).
The inverted yellow triangle marks the position of the radio
nucleus (Turner and Ho 1985).
The green line shows the approximate location 
of the dusty torus (Israel, White and Baas 1995; Dahlem et al.
in prep.).
Insets show the large-scale, smoothed images and contours overlaid.
\label{fig2}
}
\end{figure}

\clearpage

\begin{figure}
\epsscale{0.85}
\plottwo{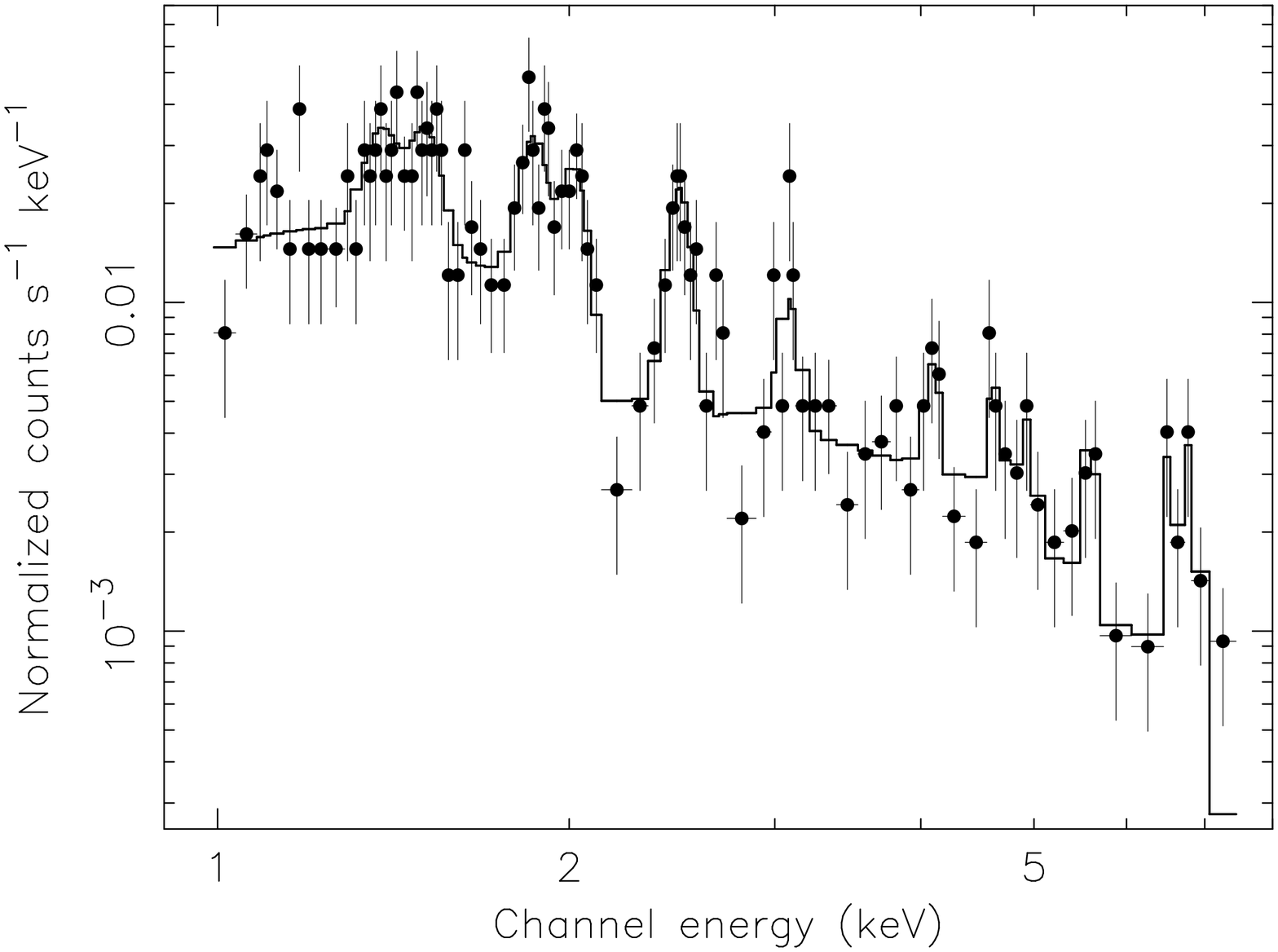}{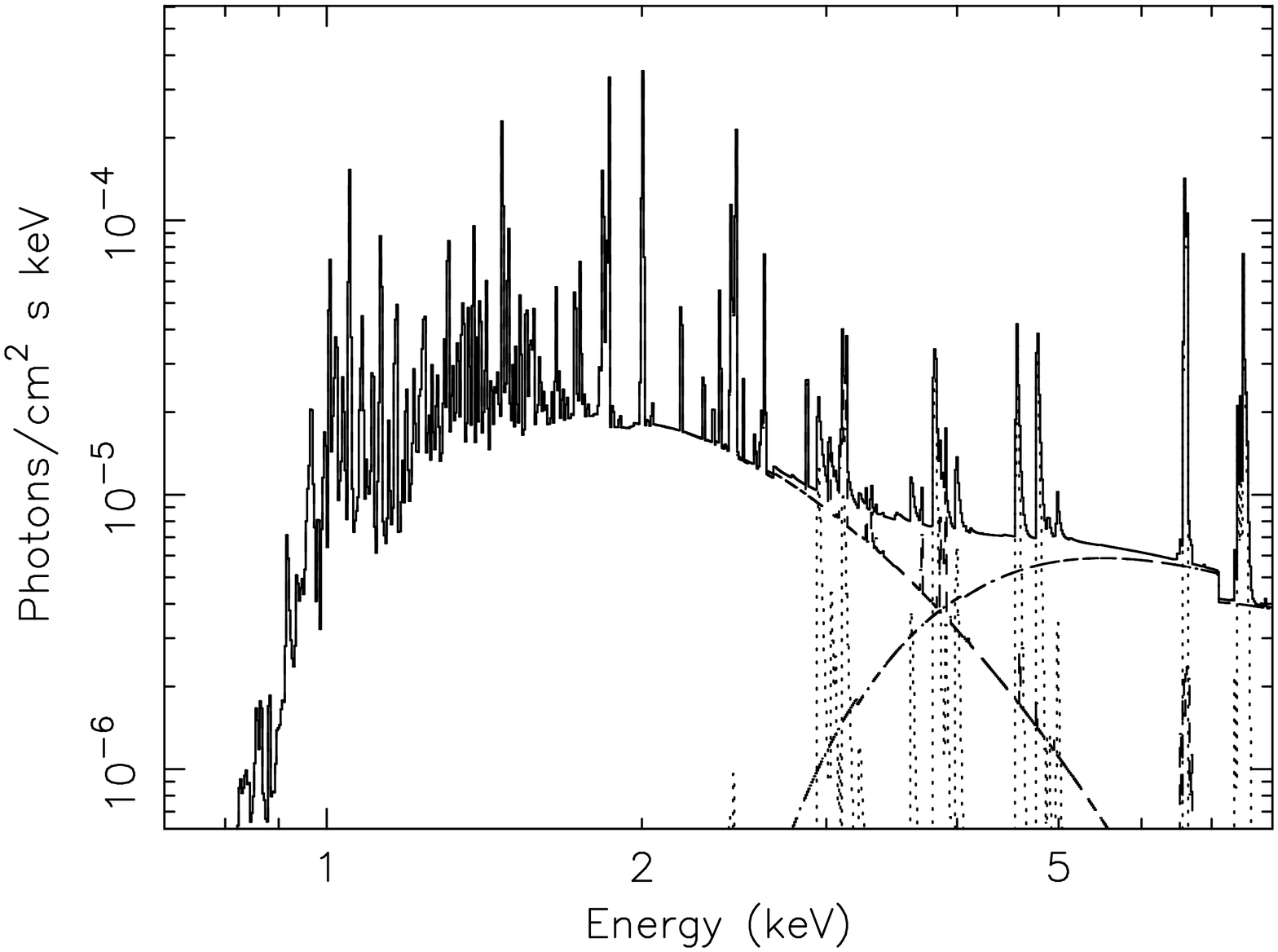}
\caption{a) The {\it Chandra} spectrum of the galaxy core and best-fitting 
empirical model (solid line) that consists of a 
power law plus 12 Gaussian emission lines (Table 1).
Data are grouped for illustration purposes such that each
energy bin contains a minimum of 5 counts. 
b) Best-fitting model for the X-ray emission at the core
of NGC~253 that consists of a low temperature thermal
plasma and a heavily absorbed hard component
($N_{\rm H}=2\times10^{23}$) that is a combination of 
a power law ($\Gamma=1.9$) plus emission from a 
photoionized plasma with $\xi\sim700$ (dotted lines).
\label{fig3}
}
\end{figure}

\begin{figure}
\epsscale{0.95}
\plotone{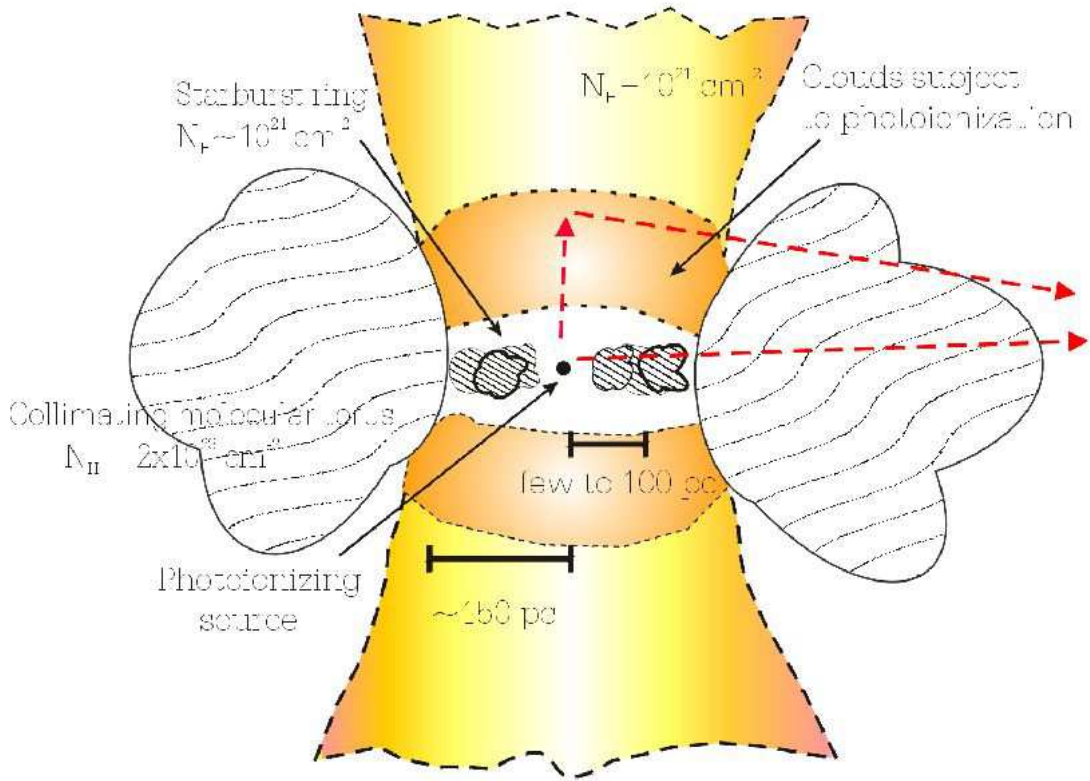}
\caption{A cartoon of the central region of NGC~253. 
\label{fig4}
}
\end{figure}

\clearpage

\begin{deluxetable}{ccccc}
\tablecolumns{5}
\tablewidth{0pc}
\tablecaption{X-ray Emission Features as Individual Gaussians \label{tab:core_fits}}
\tablehead{
Energy & norm$^a$ & EW & Luminosity & ID \\
 (keV)  & ($10^{-6}$) & (eV) & ($10^{37}$ erg s$^{-1}$) &  }
\startdata
1.40$\pm0.04$ & 3.3(1.5-5.5) & 106(50-175) & 2.0 & Mg XI \\
1.52$\pm0.03$ & 3.9(2.2-5.8) & 136(76-203) & 2.6 & Fe XXII \\
1.86$\pm0.02$ & 5.1(3.5-7.5) & 232(161-342) & 4.2 & Si XIII \\
2.02$\pm0.03$ & 4.3(2.7-6.3) & 218(136-323) & 3.7 & Si XIV \\
2.46$\pm$0.02 & 7.2(4.8-9.8) & 464(313-643) & 7.7 & S XV  \\
3.05$\pm0.05$ & 2.5(1.0-4.6) & 222(90-401) & 3.9 & S XVI - Ar XVII \\
4.08$\pm0.08$ & 1.5(0.4-3.2) & 190(56-403) & 3.2 & Ca XIX-XX \\
4.61$\pm0.08$ & 1.5(0.4-3.3) & 222(63-479) & 3.5 & Ar Rad Rec Cont ? \\
4.91$\pm0.13$ & 1.3(0.2-3.0) & 215(25-478) & 3.3 &  ? \\
5.58$\pm0.10$ & 2.2(0.6-4.2) & 410(120-800) & 5.1 & Ca Rad Rec Cont ? \\
6.50$\pm0.07$ & 3.9(1.6-7.6)  & 302(122-589) & 1.1 & Fe IV-XV \\
6.79(6.75-6.86) & 5.9(2.8-10.2) & 618(290-1,060) & 1.8 & Fe XXV   \\ 
\enddata

\tablecomments{Data are unbinned and analyzed with the c-statistic.
The continuum model is an absorbed power law with $\Gamma=1.3$, 
N$_{\rm H}=0.32\times10^{22}$ cm$^{-2}$ and a  
normalization of $4.9\times10^{-5}$ photons keV$^{-1}$ cm$^{-2}$ 
s$^{-1}$ at 1 keV.}

\tablenotetext{a}{$10^{-6}$ photons cm$^{-2}$ s$^{-1}$.
}
\end{deluxetable}

\clearpage

\begin{deluxetable}{lccccccccc}
\tablecolumns{10}
\tablewidth{0pc}
\tablecaption{Fits to Chandra Spectrum of the Nucleus of NGC~253\label{tab:fits}}
\tablehead{
\# & Model & $N_{\rm H}1^a$ & kT1$^b$ & Z / log$\xi$ & 
$N_{\rm H}2^a$ & $\Gamma$ or kT2$^b$ & 
$\chi^2$ & F$_{\rm obs}^c$ & F$_{\rm unabs}^c$ }
\startdata
1 & M   & 0.9$\pm0.3$ & 4.3$^{+3.4}_{-2.0}$ & 1f/\nix & \nix & \nix & 72/32 & 3.8 & 4.2 \\
2 & M+M & 1.8$\pm0.2$ & 1.1$\pm0.2$ & 1f /\nix & \nix & $>20.0$ & 47/30 & 4.2 & 5.0 \\ 
3 & M+M & 2.0$^{+0.2}_{-0.3}$ & 1.2$\pm0.2$ & 7.5$^{+12.0}_{-3.0}$ /\nix
     & \nix & \nix/ 20.0f & 40/29 & 4.5 & 5.5\\
4 & M+P & 2.0$^{+0.2}_{-0.3}$ & 1.2$\pm0.2$ & 1f /\nix & \nix &$-0.3^{+0.6}_{-1.5}$ /\nix 
     & 34/35 &6.8&7.9\\
5 & M+M & 2.0$^{+0.2}_{-0.3}$ & 1.2$\pm0.2$ & 1f /\nix & 45$^{+16}_{-13}$ & 
    \nix/ 1.2f & 31/30 &4.3&69.4\\
6 & M+P & 2.0$^{+0.2}_{-0.3}$ & 1.2$\pm0.2$ & 1f /\nix & 18$^{+10}_{-8}$ & 
    1.9f /\nix & 31/30 &5.1&11.4\\ 
7 & M+P & 2.0$^{+0.2}_{-0.3}$ & 1.1$^{+0.3}_{-0.1}$ & 1f / 2.6$\pm0.5$ &
  20$^{+13}_{-9}$ & 1.7f /\nix & 26/27 & 12 & 23 \\
 & +Phot$^d$ &  & &  & & &  &  & \\
\enddata
\tablecomments{Data are grouped to have at least 15 counts per bin.  Models are
M=Mekal plasma, P=Power law, Phot=Emission from cool, photoionized plasma.
f=fixed parameter.}

\tablenotetext{a}{Absorbing column density 
($N_{\rm H}$) in units of 10$^{22}$ cm$^{-2}$}
\tablenotetext{b}{kT in units of keV.}
\tablenotetext{c}{Flux is $2-10$ keV in units of 10$^{-13}$ ergs cm$^{-2}$ s$^{-1}$.
}

\end{deluxetable}

\end{document}